\input amstex.tex
\magnification\magstep1
\documentstyle{amsppt}


\define\a{\alpha}

\define\g{\gamma}

\redefine\o{\omega}

\redefine\l{\lambda}

\define\<#1,#2>{\langle #1,#2\rangle}
\define\dep(#1,#2){\text{det}_{#1}#2}
\define\norm(#1,#2){\parallel #1\parallel_{#2}}

\topmatter
\title Vacuum polarization renormalization and the geometric phase
\endtitle
\author E. Langmann and J. Mickelsson \endauthor
\affil Department of Theoretical Physics, Royal Institute of Technology,
S-10044 Stockholm, Sweden \endaffil
\endtopmatter

\document
ABSTRACT: As an application of the renormalization method introduced by the
second author we give a causal definition of the phase of the quantum
scattering matrix for fermions in external Yang-Mills potentials.
The phase is defined using parallel transport along the path of
time evolution operators. The renormalized time evolution operators are
elements of the restricted unitary group $U_{res}$ of Pressley and Segal.
The central extension of $U_{res}$ plays a central role.

\vskip 0.3in

For Dirac fermions in (non second quantized) external Yang-Mills fields
the 1-particle
scattering matrix $S$ is a well-defined unitary operator and moreover it has
a canonical second quantization $\hat S$ operating in the fermionic Fock
space [1]; for twice differentiable vector potentials one only
needs to assume certain fall-off properties at spatial and time infinity.

On the other hand, if one tries to expand the quantum scattering matrix
in ordinary Dyson-Feynman perturbation series one meets the well-known
vacuum polarization divergences, which must be taken care by suitable
(infinite) subtractions. Actually, it is only the phase of the vacuum
expectation value $<0|\hat S|0>$ which is diverging; the absolute value
is uniquely defined and finite by canonical quantization.  The crucial
point is that when passing from $S$ to $\hat S$ using the rules of canonical
hamiltonian quantization the phase remains ill-defined. It is exactly this
quantity which is diverging in perturbation theory.

The 1-particle scattering operator $S$ is defined as the limit of the
time evolution operator in the interaction picture, $U_I(t,-t) \to S$ as
$t\to\infty.$ The time evolution in the Schr\"odinger picture is defined by
$$i\partial_t U(t,t_0) = h(t) U(t,t_0) \text{ with } U(t_0,t_0)=1, \tag1$$
and in the interaction picture by
$$i\partial_t U_I(t,t_0) = V_I(t) U_I(t,t_0), \text{ with } U_I(t_0,t_0)=1.
\tag2$$
The interaction is  $V_I(t) = e^{it h_0} V(t) e^{-it h_0},$ the total
hamiltonian being $h(t)= D_A= h_0 +V(t)$ with $h_0 =D_0 = -i\g_0\g^k \partial_k.$
The quantum divergences are related to the
fact that when $V= \g^0\g^k A_k(\bold x,t) + A_0(\bold x,t)$ is the interaction
with a Yang-Mills potential then the quantization of $\hat U_I(t,-\infty)$ for
intermediate times $t < \infty$ is not well-defined. For the sake of
simplified notation we put the fermion mass $m=0.$ It does not play any role
in the treatment of ultraviolet problems below.

The starting point for our discussion here is the following renormalization
which makes $U_I$ quantizable, [2,3]. For each (smooth) potential $A=A_0 dt
+A_k dx^k$ one
defines a unitary operator $T_A$ in the 1-particle space with the following
property. Let $\epsilon= h_0/|h_0|$ be the sign of the free hamiltonian.
Define $U_{ren}(t,t_0) = T_A(t) U_I(t,t_0) T_A(t_0)^{-1}.$ Then 1)
$[\epsilon, U_{ren}(t,t_0)]$ is Hilbert-Schmidt, 2) $T_A(t) \to 1$ as $t\to
\pm \infty.$  The last property quarantees that the renormalization does not
affect the scattering matrix $S$ whereas the first condition guarantees that
the renormalized time evolution is quantizable in the free Fock space.

The Hilbert-Schmidt condition on operators $[\epsilon,g]$ defines the
restricted unitary group $U_1$ of unitary operators $g,$ [8]. The second
quantization of elements in $U_1$ defines a central extension $\hat U_1$ as
discussed in detail in [8]. Now we have a smooth path of operators $g(t)=
U_{ren}(t,-\infty)$ in $U_1$ with the initial condition $g(-\infty)=1$
and $g(+\infty)=S.$ The central extension of $U_1$ defines a natural
connection in the circle bundle $\hat U_1 \to U_1.$ The curvature of this
connection has a simple formula,          \define\ep{\epsilon}
$$\o(X,Y) = \frac14 \text{tr}\, \epsilon[\ep,X][\ep,Y]\tag3$$
where $X,Y$ are tangent vectors at $g\in U_1,$ identified as elements of the
Lie algebra of $U_1.$ The phase of $\hat S$ is defined \it through parallel
transport, \rm with respect to the connection above.

Our definition of phase of $\hat S$ is \it causal: \rm A scattering process
in an external field $A$ followed by the scattering in $A'$ defines the same
phase as the scattering for $A''=A\cup A',$ when both $A,A'$ have finite
nonoverlapping support
in time and $A''$ denotes a field $A''(t)=A(t)$ for $t <t_0$ and $A''(t)=
A'(t)$ for $t>t_0;$ here $t_0$ separates the supports of $A,A'.$
                               \define\Asl{A\!\!\!\!\slash}
The operator $T_A$ is not uniquely defined. It is more convenient to define
the transformation first in the Schr\"odinger picture (1).
A simple formula which works in the gauge $A_0=0$ is (here $E\!\!\!\!\slash
=\partial_t \Asl$)
$$T_A = \exp\left(\frac14 [\frac{1}{D_0}, \Asl] -
\frac18[\frac{1}{D_0}\Asl\frac{1}{D_0},\Asl ]-\frac{i}{4} \frac{1}{D_0}
E\!\!\!\!\slash \frac{1}{D_0} \right), \tag4  $$
where it is understood that the singularity at the zero modes of $D_0$ is taken
care of by an infrared regularization, for example $\frac{1}{D_0} \to
\frac{D_0}{D_0^2 + \a^2}$ for some nonzero real number $\a.$ In the interaction
picture one uses the operator $\exp(ith_0) T_A \exp(-ith_0).$

The validity of the choice (4) is proven as follows. The time evolution equation
for $U_{ren}(t,-\infty) = T_A U(t,-\infty)$ is
$$i\partial_t U_{ren}(t) = (h_0 +W(t)) U_{ren}(t)$$
with
$$W(t) = (i\partial_t T_A) T_A^{-1}  +T_A (h+V(t)) T_A^{-1}.$$
Expanding the exponential and arranging terms according to powers of the
inverse of momentum (i.e., of $D_0$,) one gets $[\ep, W]= R_1 +R_2 +\dots,$ where
the dots denote terms which behave explicitly as $|D_0|^k$ with $k\leq -2$
for high momenta, and                         \define\DD{\frac{D_0}{|D_0|}}
$$R_1= \frac12\frac{D_0}{|D_0|}\Asl -\frac12 \Asl \DD +\frac14 |D_0|\Asl\frac{1}
{D_0} -\frac14 D_0 \Asl\frac{1}{|D_0|} +\frac14 \frac{1}{|D_0|} \Asl D_0
-\frac14 \frac{1}{D_0}\Asl |D_0|       $$
and the second term, which is quadratic in $\Asl$, is \define\DI{\frac{1}{D_0}}
$$\align R_2 =&\frac14 \DI [|D_0|,\Asl^2] \DI +\frac14 \left[\frac{1}{|D_0|},
\Asl^2\right]
+\frac18 D_0 \left[\Asl\DI\Asl,\frac{1}{|D_0|}\right] +\\ &+\frac18
\left[\Asl\DI\Asl, \frac{1}{|D_0|}\right] D_0
+\frac18 \DI \left[\Asl\DI\Asl, |D_0|\right] +\frac18 \left[\DI\Asl\DI,
|D_0|\right] \DI.\endalign$$
Since the commutator $[|D_0|^k,\Asl]$ is of order $|D_0|^{k-1}$ in momenta,
all terms in $R_2$ are actually of order $-2$ or less. In order to estimate
$R_1$ we write it in equivalent form
$$R_1 = \frac14 \DD \left[ [\Asl, |D_0|], \frac{1}{|D_0|} \right]
-\frac14 \left[ [\Asl, |D_0|], \frac{1}{|D_0|}\right] \DD$$
and observe both terms are of order $-2$ for the same reason as in the case
of $R_2.$ We have disregarded all the low order terms because in three space
dimensions (in a box) the condition that a PSDO is Hilbert-Schmidt (which was
required for canonical quantization) is precisely the requirement that the
operator vanishes for high momenta faster than $|p|=|D_0|$ raised to power
$-3/2.$ Thus after our renormalization (= conjugation by the time-dependent
unitary operator $T_A$) the gauge interaction can be lifted to a finite
operator in the fermionic Fock space.

This method can be extended to other interactions as well, under very
general conditions [3]. It has been also used to derive the chiral anomaly
in the hamiltonian framework [2], [3].

The curvature formula  (3) is \it nonlocal. \rm The computation of the
trace involves space derivatives up to arbitrary high power because of the
Green's function $1/D_0$ in the renormalization prescription. However, the
curvature is equivalent in cohomology to a local formula. For bounded
pseudodifferential operators satisfying the condition that the degree of the
commutator $[\ep,X]$ is less than $-n/2$ (here the space dimension $n=3$) we
have
$$\o_{loc}(X,Y)= \text{Res}\,\ep [\text{log} |p|,X] Y,\tag5$$
where $|p|,$ the length of three momentum, is the symbol of $|D_0|.$
The difference $\o -\o_{loc}$ is a coboundary, [2], [4]. Here we have used
the operator residue [7] for PSDO's.
Without the sign $\ep$ this formula would give the Radul cocycle, which is a
2-cocycle on the algebra of \it all \rm PSDO's on a compact manifold [5]. In
quantum field theory it is important to keep the sign because this is the
cocycle arising from normal ordering. Let us recall how the normal ordering is
related to the curvature $\o$ and Schwinger terms.

We want to concentrate on the ultraviolet
behavior of operators and therefore we put the system in a box (or assume from
very beginning that the Dirac particle is moving on a compact manifold).
Let $a^*_i, a_j,$ with $i,j\in\Bbb Z,$ be a complete set of
creation and annihilation operators such that the index $i\geq 0$ corresponds
to nonnegative energies and $i<0$ to negative energies,
$$a^*_i a_j +a_j a^*_i = \delta_{ij}.$$
The normal ordering is defined by $:a^*_i a_j : = a^*_i a_j$ except when $i=j
<0$ and then $:a^*_i a_j: = - a_j a^*_i.$
If $X,Y$ are a pair of one-particle
operators then the canonical quantizations
$$\hat X = \sum X_{ij}  : a^*_i a_j : $$
satisfy the algebra
$$[\hat X,\hat Y] = \widehat{[X,Y]} +\o(X,Y),$$
and the quantum operators are defined such that $[\hat X, a^*_i] =\sum X_{ji}
a^*_j .$ The equivalence of the two cocycles means simply that if $\tilde X=
\hat X +\l(X)$ for a suitable complex linear form on the algebra of one-particle
operators then
$$[\tilde X, \tilde Y] = \widetilde{[X,Y]} + \o_{loc}(X,Y).\tag6$$
It is also interesting to note that the local 2-cocycle can be written as
$$\o_{loc}(X,Y) =\text{Res} \frac{1}{|D_0|} [D_0,X] Y.\tag7$$
This is one of the cyclic cocyles in noncommutative geometry [6]. (But
remember that in three space dimensions we must impose the restrictions
mentioned earlier on the PSDO's $X,Y.$)

The great advantage of the residue formula is that it only depends on the term
in
the appropriate PSDO which is precisely of asymptotic degree $-n$ (here $-3$)
when expanded asymptotically in powers of $1/|p|.$ In operator products each
derivative in configuration space is associated with a derivative in momentum
space. Since a differentiation in momentum space decreases the homogeneous
degree of the operator it follows that a calculation of the residue can
involve derivatives with respect to $x_i$'s only up to a finite order.

We finally comment on the renormalization dependence of the phase of the
quantum scattering matrix $\hat S.$
Suppose that we have constructed another renormalization operator $T'_A$ such
that  each term in the asymptotic expansion
$$T'_A = 1 + t_{-1}(A) + t_{-2}(A) +\dots,$$
in homogeneous terms $t_k$ of degree $k$ in momenta, is a local differential
polynomial in the components of the vector potential $A.$ Then the change of
phase of the quantum scattering is given by a parallel transport around
a loop of time evolution operators obtained by first travelling the path
$g(t)$ from the point $g(-\infty)=1$ to $g(\infty)=S$ defined by the renormalized
time evolution due to $T_A$ and then following backwards in time the time
evolution defined by the renormalization $T'_A.$  But the logarithm of the
parallel transport phase is equal to the integral of the curvature over
a surface bounded by the loop. Since the curvature formula is local,
the value of the integral is given by a space integral of a local differential
expression in the vector potential. This is in the spirit of local quantum
field theory: A change in the renormalization corresponds to a local counterterm
in the Lagrangian.

Generically, a given choice of $T_A$ does not lead to a gauge invariant
phase even in the case of Dirac fermions. (For Weyl fermions this would be
expected because of the chiral anomaly.) But this is not a serious problem
and it can be avoided by introducing local counterterms to the hamiltonian
as will be discussed in detail in [9].

\vskip 0.2in
\bf References \rm \newline\newline
\noindent [1] J. Palmer,
J. Math. Anal. Appl \bf 64, \rm 189 (1978). S.N.M Ruijsenaars, J. Math Phys.
\bf 18, \rm 720 (1976) and J. Funct. Anal. \bf 33, \rm 47 (1979)\newline\newline
[2] J. Mickelsson, In:
\it Integrable Systems and Strings. \rm
Ed. by A. Alekseev et al. Springer Lecture Notes in Physics \bf 436 \rm (1994)
\newline\newline
[3] E. Langmann and J. Mickelsson,
J. Math. Phys. \bf 37, \rm  3933 (1996)                  \newline\newline
[4] M. Cederwall, G. Ferretti, B.E.W. Nilsson, A. Westerberg, Commun. Math.
Phys. \bf 175, \rm 221 (1996)       \newline\newline
[5] A.O. Radul, Funct. Anal. Appl. \bf 25, \rm 33 (1991)
\newline\newline
[6] A. Connes: \it Noncommutative geometry. \rm
Academic Press (1994)   \newline\newline
[7] M. Wodzicki,  In: \it
K-theory, arithmetic and geometry. \rm (ed. by Yu.I. Manin)
Lect. Notes in Math. \bf 1289, \rm
320-399, Springer-Verlag, Berlin, Heidelberg, New York (1987)\newline\newline
[8] A. Pressley and G. Segal: \it Loop Groups. \rm Clarendon Press, Oxford
(1986)\newline\newline
[9] J. Mickelsson, in preparation

\enddocument